\documentclass[pre,amssymb,amsmath,floats,twocolumn,showpacs,superscriptaddress]{revtex4}

\usepackage{epsfig}

\begin{document}

\title{Degree-dependent intervertex separation in complex networks} 

\author{S. N. Dorogovtsev}
\email{sdorogov@fis.ua.pt}
\affiliation{Departamento de F{\'\i}sica da Universidade de Aveiro, 3810-193 Aveiro, Portugal}
\affiliation{A. F. Ioffe Physico-Technical Institute, 194021
  St. Petersburg, Russia} 

\author{J. F. F. Mendes}
\email{jfmendes@fis.ua.pt}
\affiliation{Departamento de F{\'\i}sica da Universidade de Aveiro, 3810-193 Aveiro, Portugal}

\author{J. G. Oliveira}
\email{joaogo@fis.ua.pt}
\affiliation{Departamento de F{\'\i}sica da Universidade de Aveiro, 3810-193 Aveiro, Portugal}

%%\date{\today}
\date{}

\begin{abstract} 
We study the mean length $\ell(k)$ of the shortest paths between a vertex of degree $k$ and other vertices in growing networks, where correlations are essential. In a number of deterministic scale-free networks we observe a power-law correction to a logarithmic dependence, 
$\ell(k) = A\ln [N/k^{(\gamma-1)/2}] - C k^{\gamma-1}/N + \ldots$ in a wide range of network sizes. 
%%%%%%**********  
%%%%%%We find the mean length $\ell(k)$ of the shortest paths between a vertex of degree $k$ and other vertices in growing networks. In scale-free networks, we obtain a power-law correction to a logarithmic dependence, 
%%%%%%$\ell(k) = A\ln [N/k^{(\gamma-1)/2}] - C k^{\gamma-1}/N + \ldots$. 
%%%%%%%%$\ell(k) = A\ln N - B\ln k - C k^{\gamma-1}/N + \ldots$. 
%%%%%%%%A \ln (N/k)/\ln \overline{k} + B k^{\gamma-1}/N$. 
Here $N$ is the number of vertices in the network, $\gamma$ is the degree distribution exponent, 
%%$P(k) \propto k^{-\gamma}$, 
and the coefficients  
%%$\overline{k}$ is its mean degree, 
$A$ 
%%, $B$, 
and 
$C$ depend on a network. 
%%%%%%%%%%In random scale-free networks, growing through the preferential attachment mechanism, at sufficiently low degrees, we observe the following form: $\ell(k) \approx  A\ln N - B\ln k + \ldots$, where $A \neq B$. 
We compare this law with 
%%the 
a 
corresponding $\ell(k)$ dependence  
%%%%%%%%%%more complex one 
obtained for random scale-free networks growing through the preferential attachment mechanism. 
%%%%%We obtain this result for a number of growing deterministic graphs but believe that it holds for a wide class of evolving scale-free networks. 
%%%%%%%%%%Contrastingly, in 
In stochastic and deterministic growing trees with an exponential degree distribution, we 
%%%%%%find 
observe 
a linear dependence on degree, $\ell(k) \cong A\ln N - C k$. 
We compare our 
%%%%%results 
findings 
for growing networks with those for uncorrelated graphs. 

\end{abstract}

\pacs{05.50.+q, 05.10.-a, 05.40.-a, 87.18.Sn}

\maketitle

\section{Introduction}\label{s-introduction} 

%%${\cal T}$

The main objects of interest of the physics of complex networks \cite{ab02,dm02,n03,wbook99,dmbook03,pvbook04} are extremely compact, infinite dimensional nets---so called small worlds. 
The basic measure of the compactness of a network is the mean intervertex distance or the mean intervertex separation, that is, the mean length of the shortest path between a pair of vertices, $\ell$. (The path runs along edges, each edge has the unit length.) Physicists often use another term for this characteristic---the diameter of a network, although in graph theory the term network diameter is reserved for the maximal separation of a pair of vertices in a net. 

A network shows the small-world effect if its mean intervertex distance slowly increases with the network size (the total number of vertices in a network, $N$), slower than any power-law function of $N$. 
This is in contrast to finite dimensional objects, where the mean intervertex distance grows as $N^{1/d}$, $d$ being the dimension of an object. (We discuss sparse networks.)  
By definition, a small world is a network with the small-world effect. 
Note that this definition is not related to the presence of loops in 
a network. 
Small worlds may be loopy or clustered networks, or they may be without loops---trees. 

The mean intervertex distances in networks were extensively studied both in the framework of empirical research \cite{ajb02} and analytically \cite{nsw02,dms03,ch02,ffh02}. 
The typical size dependence of the mean intervertex separation is logarithmic, $\ell(N) \propto \ln N$. However, the mean intervertex distance is an integrated, coarse characteristic. 
One may be interested in a more delicate issue---the position of an individual vertex in a network. 
%%%%%One should note that recently 
Recently Holyst {\it et al.} 
%%, Ref. 
\cite{ffh04}, have considered the question: 
how far are vertices of specific degrees from each other? 
%%******** ***** *****  
They have shown that in uncorrelated networks, the mean length of the shortest path between vertices of degrees $k$ and $k'$ is 
$\ell(k,k') \cong D + A\ln N - A\ln(kk')$, where $D$ is independent of $N$, $k$, and $k'$, and the coefficient $A$ depends only of the mean branching ratio of the network. 
Note the coincidence of the coefficients of $\ln N$ and $\ln(k,k')$ in this result. 
The authors of paper~\cite{ffh04}, also calculated $\ell(k,k')$ of networks with nonzero clustering though without degree-degree correlations. In this case, they have arrived at the same expression as above but with 
%%$A$ and $B$ additionally depending on the clustering coefficient.  
coefficients of $\ln N$ and $\ln(k,k')$ additionally depending on the clustering. 
%%******* ***** ***** 
In the present paper we present our observations for another (though related) characteristic---the mean length of the shortest paths from a vertex of a given degree $k$ to the remaining vertices of the network, $\ell(k)$. This quantity is related to $\ell(k,k')$ in the following way:   
%%. 
%%
\begin{equation} 
\ell(k) = \sum_{k'} P(k')\ell(k,k')
\, , 
\label{c1}
\end{equation} 
and so  
\begin{equation} 
\ell = \sum_k P(k)\ell(k) = \sum_{k,k'} P(k)P(k') \ell(k,k')
\, .  
\label{c2}
\end{equation} 
In simple terms, we reveal the smallness of a network from the point of view of its vertex of a 
given degree. Our objects of interest are growing (and so inevitably correlated) networks.  
%%%%%%of a complex network 

The basic property of 
%%the 
most of the natural networks is a heavy-tailed degree distribution, so that vertex degrees are distributed over a wide range in contrast to classical random graphs. This motivates the study of $\ell(k)$ in networks with various complex distributions of connections. The question is: How strong the variation of $\ell(k)$ may be?  
One should note that this characteristic was measured recently in Ref.~\cite{mk04} in several networks, and noticeable variations of $\ell(k)$ were found. 
We 
%%%%%obtain analytically 
observe nontrivial dependences $\ell(k)$ for networks with power-law and exponential degree distributions. We mostly consider growing networks, where correlations between the degrees of vertices are important, but for comparison, also discuss uncorrelated networks. 
In our study we use convenient deterministic growing graphs and compare 
%%%%%some of our results 
our observations with simulations of stochastic models of growing networks. 
%%%%%%%%%%%%It is important that at sufficiently high $k$, we observe the law $\ell(k) = D + A\ln N - B\ln k + \ldots$, where the coefficients $A$ and $B$ are essentially different. 

In Sec.~\ref{s-results} we list our 
main 
%%%%%results, 
observations, so that readers not interested in details may restrict themselves to the first two sections. Section~\ref{s-uncorrelated} contains the discussion of the $\ell(k)$ dependence in uncorrelated networks for the sake of comparison. 
%%Section~\ref{s-derivations} explains 
In Sec.~\ref{s-derivations} we explain 
in detail how the results were obtained and describe particular cases. 
In Sec.~\ref{s-discussion} we make a few remarks on the degree-dependent intervertex separation in various networks and discuss relations of this quantity to centrality measures used in sociology \cite{s66,n90}.

%%%%%\section{Main results}\label{s-results} 

\section{Main observations}\label{s-results}

%%%%%%%%%%%%%%%%%%%%%%%%%%%%%%%%%%%%%%%%%%%%%%%%%%%%%%%%%%%%%%%%%%%%
%%%%%%%%%%%%%%%%%%%%%%%%%%%%%%%%%%%%%%%%%%%%%%%%%%%%%%%%%%%%%%%%%%%%

\begin{figure*}%%[h]
\epsfxsize=140mm
%%\epsffile{degree-intervertex_fig1.eps}
\epsffile{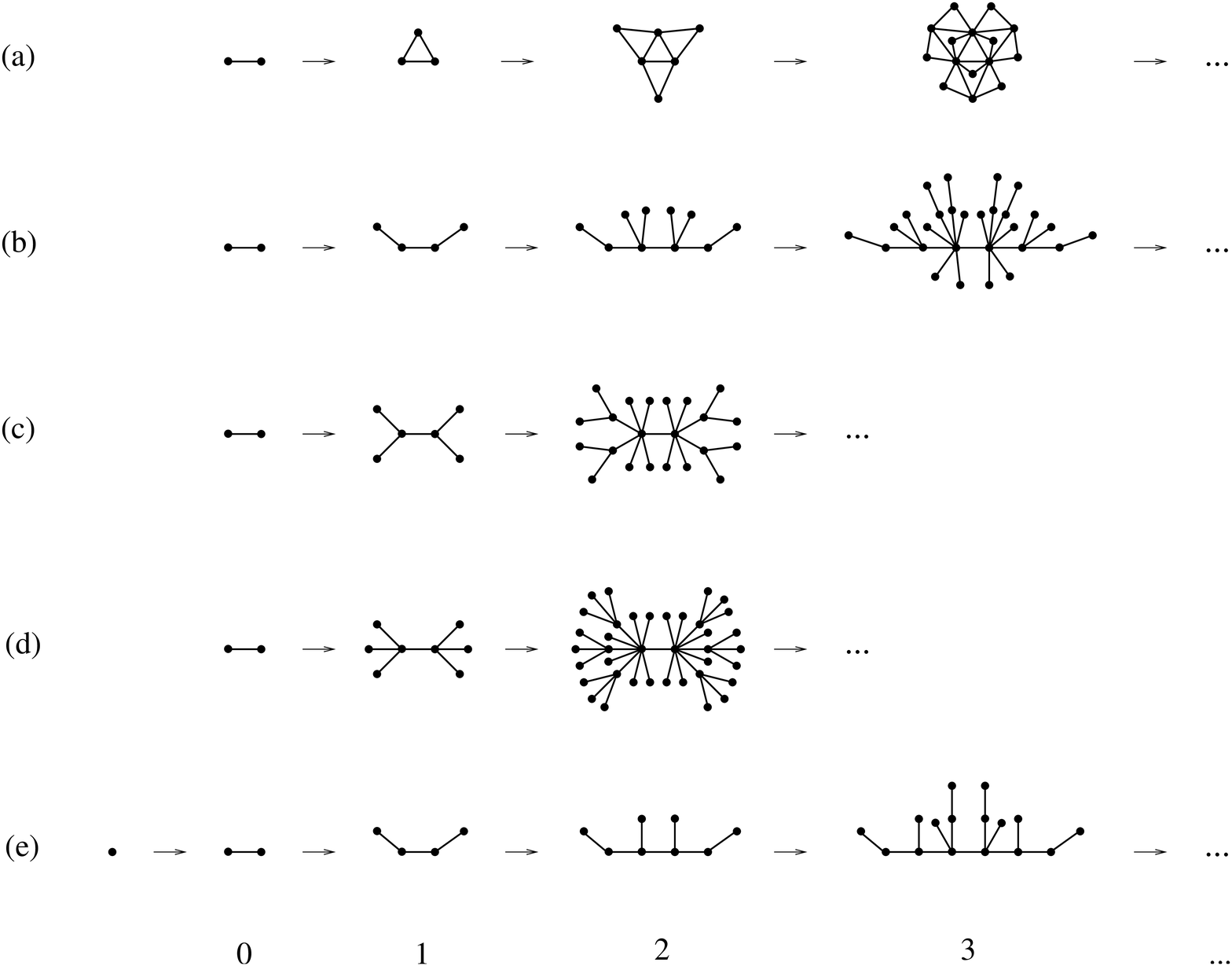}
\caption{ 
The set of deterministic graphs that is used in this paper. 
(a) A scale-free graph with the exponent of the degree distribution $\gamma=1+\ln3/\ln2 = 2.585\ldots$ \protect\cite{dm02,dgm02}. 
At each step, each edge of the graph transforms into a triangle. 
(b) A scale-free tree graph with $\gamma=1+\ln3/\ln2 = 2.585\ldots$ 
\protect\cite{jkk02}. 
At each step, a pair of new vertices is attached to the ends of each edge of the graph.  
(c) A scale-free tree graph with $\gamma=3$. 
At each step, a pair of new vertices is attached to the ends of each edge plus 
a new vertex is attached to each vertex of the graph.  
(d) A scale-free tree graph with $\gamma = 1+\ln5/\ln2 = 3.322\ldots$. 
At each step, a pair of new vertices is attached to the ends of each edge plus 
two new vertices are attached to each vertex of the graph.
(e) A deterministic graph with an exponentially decreasing spectrum of degrees \protect\cite{jkk02}. 
At each step, a new vertex is attached to each vertex of the graph. 
In all these graphs, a mean intervertex distance grows with the number $N$ of vertices as $\ln N$.
}
\label{f1}
\end{figure*}

%%%%%%%%%%%%%%%%%%%%%%%%%%%%%%%%%%%%%%%%%%%%%%%%%%%%%%%%%%%%%%%%%%%%
%%%%%%%%%%%%%%%%%%%%%%%%%%%%%%%%%%%%%%%%%%%%%%%%%%%%%%%%%%%%%%%%%%%% 

For the purpose of the analytical description of $\ell(k)$ we use  
%%%%%Our results were obtained by using 
simple deterministic graphs. 
%%%%%which allow exact solution of the problem. 
Deterministic small worlds were considered in a number of recent papers \cite{brv01,dgm02,jkk02,noh03,nr04,cfr04,rkba04,ahas04,dma04,dm04,ba04} and have turned out to be a useful tool. 
(We called these networks {\em pseudofractals}. 
Indeed, at first sight, they look as fractals. However, they are infinite dimensional objects, 
so that they are not fractals.)
These graphs correctly reproduce practically all known network characteristics. We use a set of deterministic scale-free models with various values of the degree distribution exponent $\gamma$, $P(k) \propto k^{-\gamma}$ (see Fig.~\ref{f1}). 
We consider deterministic graphs with $\gamma$ in the range between $2$ and $\infty$, where a graph with $\gamma=\infty$ has an exponentially decreasing (discrete) spectrum of degrees. 

In the studied scale-free deterministic graphs, in a wide range of the graph sizes, the 
%%%%%result 
%%observation for the 
mean separation of a vertex of degree $k$ from the remaining vertices of the network is found to follow the dependence: 
%%%%%looks as follows:  
%%
\begin{equation} 
\ell(k) = 
%%A\ln[N/k^{(\gamma-1)/2}] - C k^{\gamma-1}/N + \ldots 
A\,\ln\!\left[\frac{N}{k^{(\gamma-1)/2}}\right] - C\, \frac{k^{\gamma-1}}{N} + \ldots 
\, .
\label{e1}
\end{equation} 
%%%$\phantom{.}$
%%%\noindent
The constants $A$ and $C$ (as well as the sign of $C$) depend on a particular network. 
%%%%%%%%%%%%%%%This law  
%%%%%asymptotic formula is 
%%%%%%%%%%%%%%%is observed for large enough $k$. 

%%%%%%%%%%%%%%%%%%%%%%%%%%%%%%%%%%%%%%%%%%%%%%%%%%%%%%%%%%%%%%%%%%%%
%%%%%%%%%%%%%%%%%%%%%%%%%%%%%%%%%%%%%%%%%%%%%%%%%%%%%%%%%%%%%%%%%%%%

\begin{figure}
\epsfxsize=61mm
%%\epsffile{degree-intervertex_fig2.eps}
\begin{center}
%\scalebox{0.28}{\includegraphics[angle=270]{degree-intervertex_fig2prime-prime.ps}}
%%\scalebox{0.28}{\includegraphics[angle=270]{degree-intervertex_fig2a.ps}}
%%\scalebox{0.28}{\includegraphics[angle=270]{degree-intervertex_fig2b.ps}}
\scalebox{0.28}{\includegraphics[angle=270]{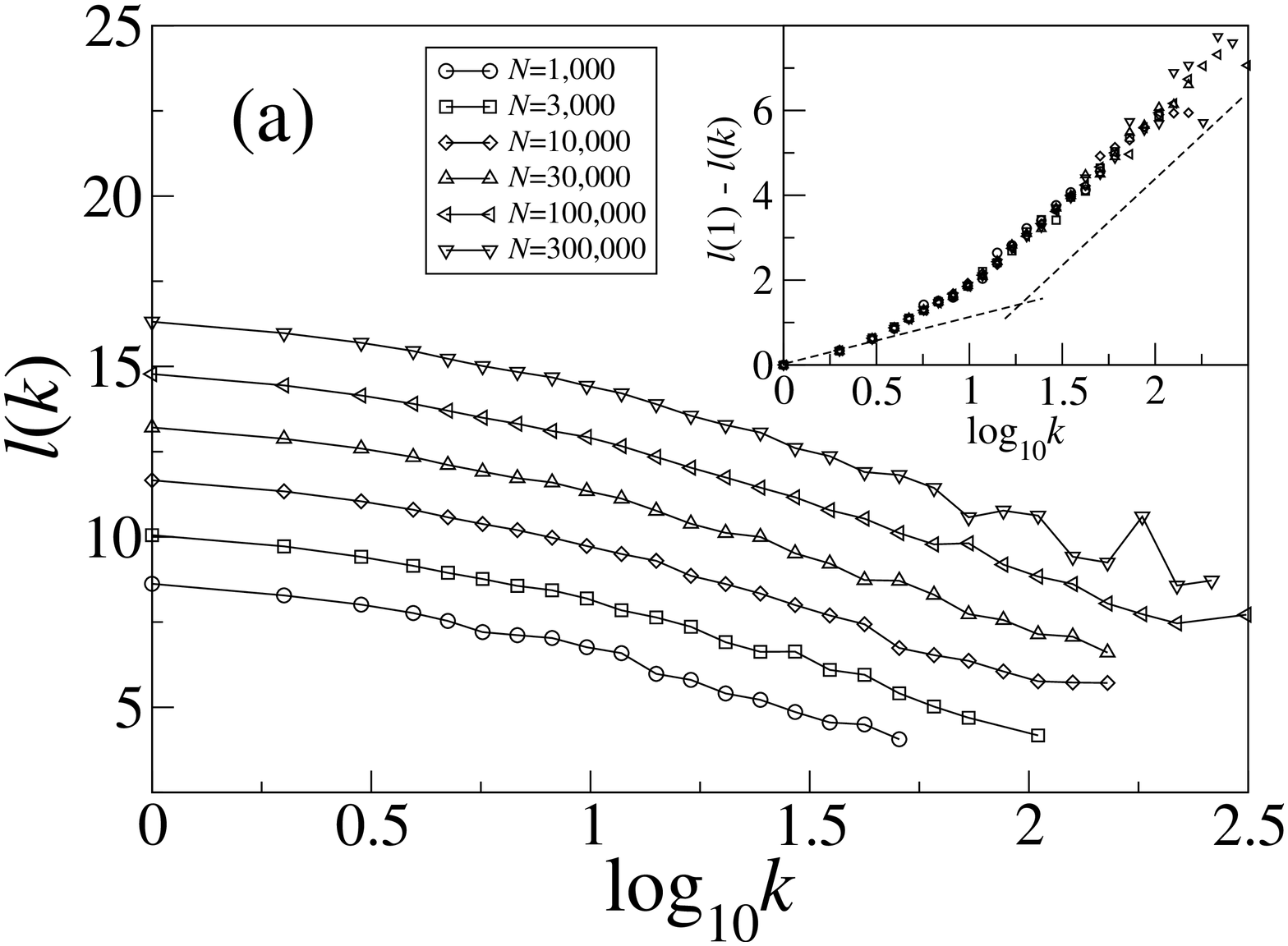}} 
\scalebox{0.28}{\includegraphics[angle=270]{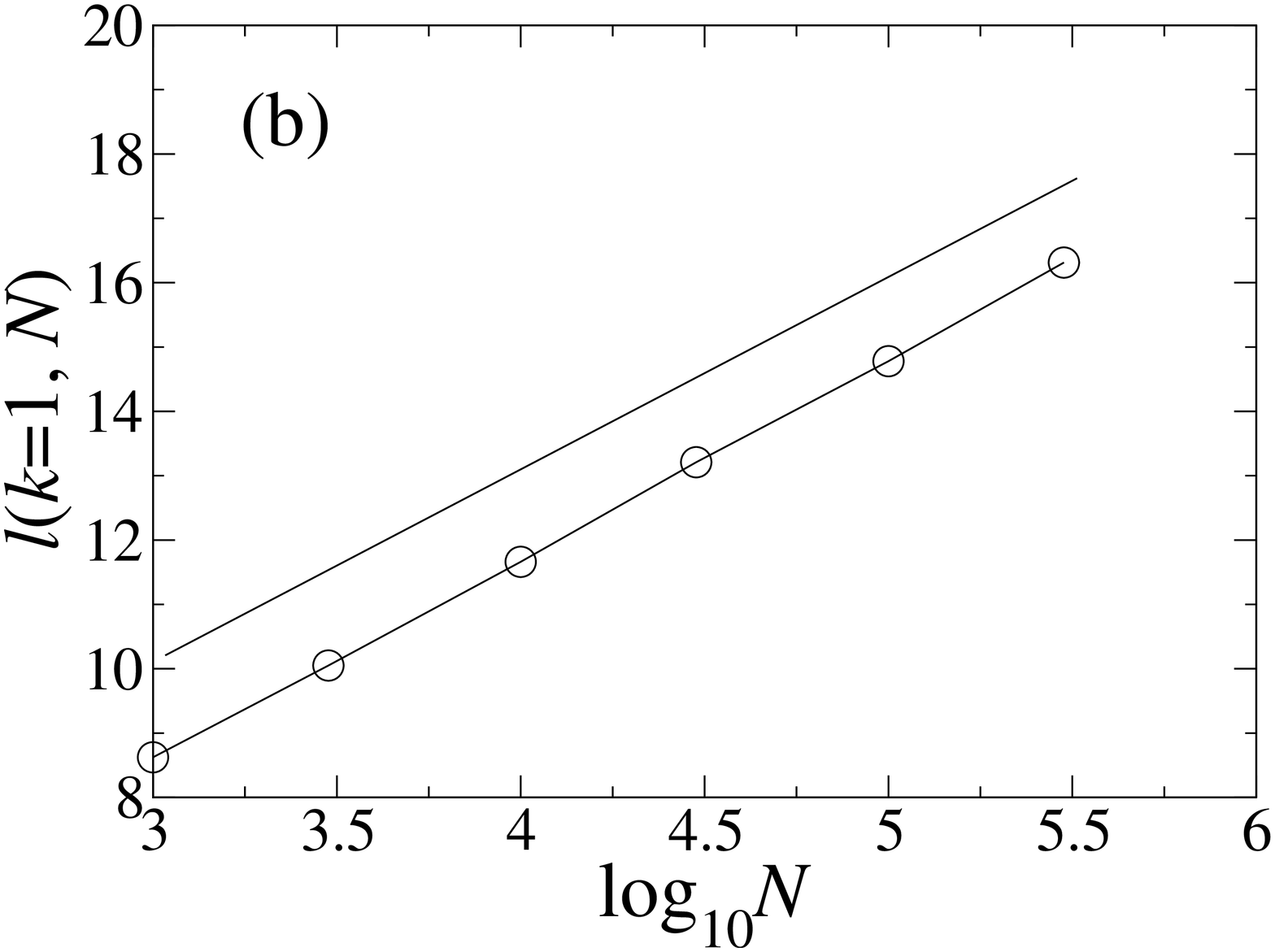}} 
\end{center}
%%\epsffile{degree-intervertex_fig2prime-prime.ps}
\caption{
Degree-dependent mean intervertex separation in a random scale-free network (tree) growing through the mechanism of preferential attachment. 
At each time step a new vertex is added. It becomes attached to a vertex selected with probability proportional to the sum of the degree of this vertex and a constant $A$ --- ``additional attractiveness'' \protect\cite{dms00}. Here we use $A=1$.  (a) $\ell(k)$ vs. $\log_{10}k$ for networks of $N=1000,$ $3000,$ $10\,000,$ $30\,000,$ $100\,000,$ and $300\,000,$ vertices. Each of the first four curves were obtained 
%%%by using 
after $50$ runs, while for the networks of $100\,000$ and  $300\,000$ vertices, $20$ and $5$ runs were used, correspondingly. 
Binning was made at large degrees, which allowed us to reduce noise. 
The inset demonstrates that in this network, the difference $\ell(k=1)-\ell(k)$ does not depend on the size $N$. In the inset, for the sake of clearness we do not show lines connecting points. The dashed lines highlight two limiting behaviors. 
%%---at small and at large degrees.  
As $k$ approaches its minimal value $k=1$, $\ell(k=1)-\ell(k) \approx 1.0\, \log_{10}k 
\approx 0.43\, \ln k$ 
for all studied network sizes, while at large degrees, $\ell(k=1)-\ell(k) \approx \text{const} + 4.1\, \log_{10}k 
\approx \text{const} + 1.8\, \ln k$. 
(b) The dependence of $\ell(k=1)$ on $\log_{10}N$. For comparison, a line with a slope $3$ is shown. 
%%This slope differs from the slopes of both dashed lines in the inset above.   
%%%These networks have exponential degree distributions. 
%%%(a) At each time step, a vertex is attached to a randomly chosen vertex of the network. The dependence is the result of the simulation of the network of $10^5$ vertices, $50$ runs. For comparison, a line with a slope $-1/2$ is shown.  
%%%(b) At each time step, 3 vertices are attached to a randomly chosen vertex of the network. The dependence is presented for the network of $9998$ vertices, $50$ runs. 
%%%The initial configuration consists of two vertices connected by an edge. 
%%%For comparison, a line with a slope $-1/4$ is shown. 
%%%Note that in these plots $\max \ell(k) \approx 2\min  \ell(k)$. In other words, in these networks, there are no vertices of degree greater than $k_\text{max}$: $\ell(k_\text{max})=\max \ell(k)/2$. Note fluctuations in the range of the highest degrees.  
}
\label{f1prime}
\end{figure}

%%%%%%%%%%%%%%%%%%%%%%%%%%%%%%%%%%%%%%%%%%%%%%%%%%%%%%%%%%%%%%%%%%%%
%%%%%%%%%%%%%%%%%%%%%%%%%%%%%%%%%%%%%%%%%%%%%%%%%%%%%%%%%%%%%%%%%%%%

In stochastic growing scale-free networks, we 
%%also 
observe a dependence $\ell(k,N)$ shown in Figure~\ref{f1prime}. 
%%similar 
%%%%%%%%%%%%%%%%%dependence $\ell(k)$ with distinct coefficients of $\ln N$ and $\ln k$. 
%%%%%%%%%%%%%%%%%Figure~\ref{f1prime} 
This figure demonstrates the results of the simulations of networks growing by the preferential attachment mechanism with a linear preference function \cite{dms00}. 
While the dependence on $\ln N$ is linear practically in the entire range of observation, $\ell(k)$ vs. $\ln k$ is of a more complex form (see Fig.~\ref{f1prime}). The derivative $d\ell(k)/d\ln k$ is non-zero at $k=1$ and at large degrees, $\ell(k)$ is fitted by a linear function of $\ln k$ with a larger slope.   
%%We believe that this dependence is valid for a wide class of random, growing, scale-free networks. 
One should note that in all growing networks considered in this paper, new connections cannot emerge between already existing vertices. These networks are often called ``citation graphs''. 
%%
%%%%%For more general scale-free graphs, one may suggest the main contribution of the form $\ell(k) \cong A \ln N - B \ln k$.  

In the specific point $\gamma=3$, correlations between the degrees of the nearest neighbors in these graphs are anomalously low. In this situation, the main contribution to $\ell(k)$ reduces to $\ell(k) \propto \ln(N/k)$, which coincides with the result for equilibrium uncorrelated networks (see the next section). 

\begin{figure}
\epsfxsize=61mm
%%\epsffile{degree-intervertex_fig2.eps}
\begin{center}
%\scalebox{0.28}{\includegraphics[angle=270]{degree-intervertex_fig2prime-prime.ps}}
%%\scalebox{0.28}{\includegraphics[angle=270]{degree-intervertex_fig2a.ps}}
%%\scalebox{0.28}{\includegraphics[angle=270]{degree-intervertex_fig2b.ps}}
\scalebox{0.28}{\includegraphics[angle=270]{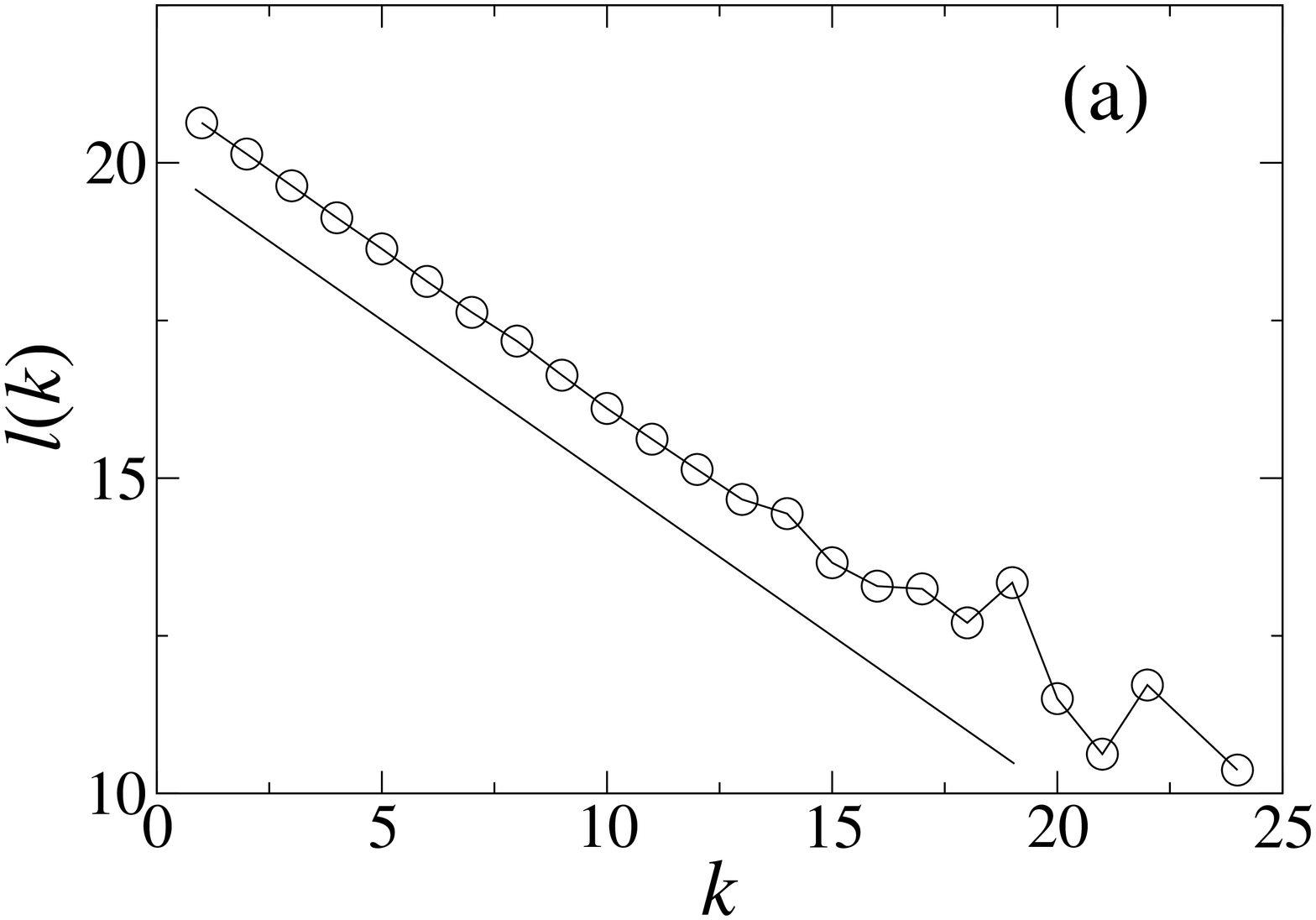}}
\scalebox{0.28}{\includegraphics[angle=270]{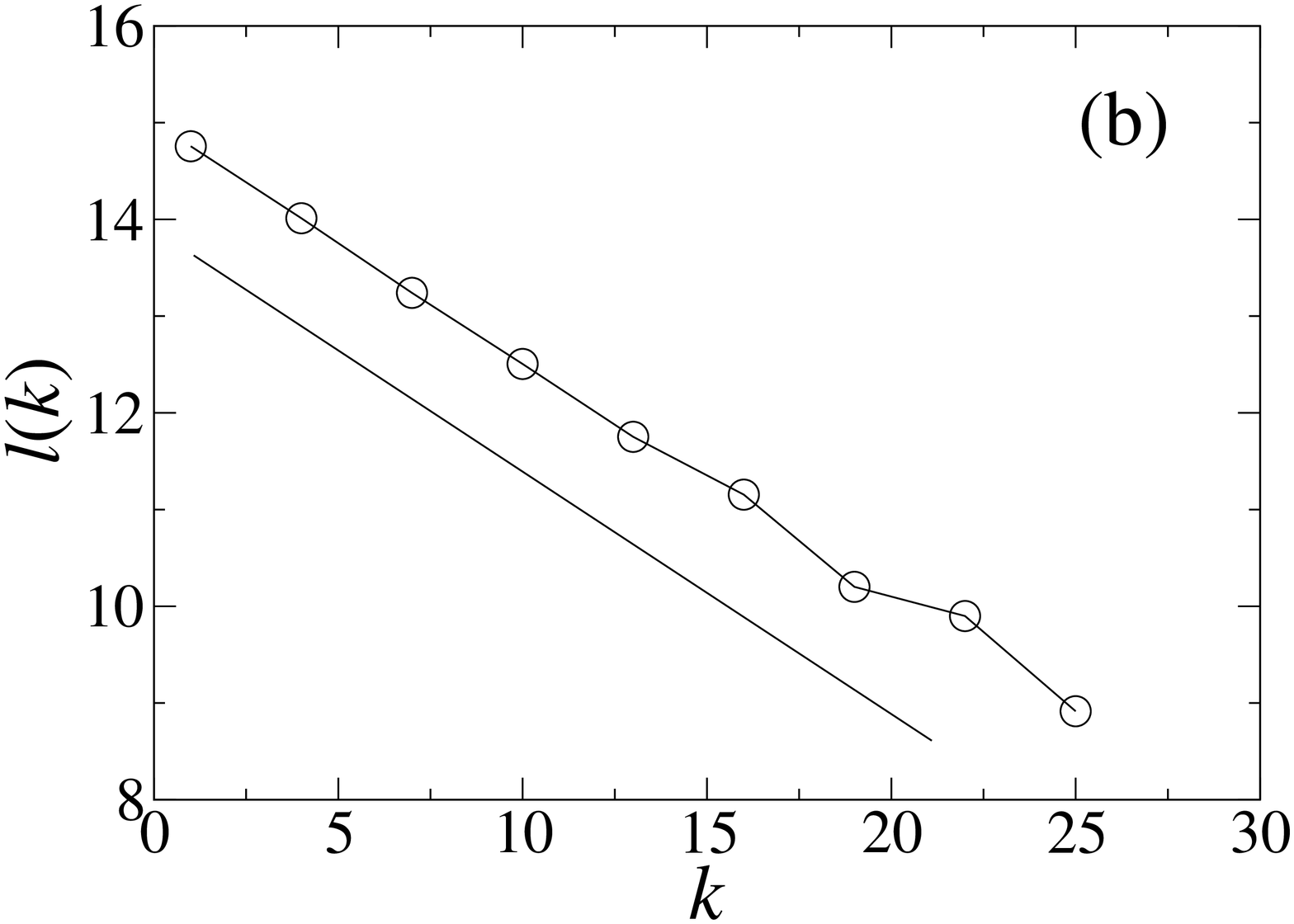}}
 \end{center}
%%\epsffile{degree-intervertex_fig2prime-prime.ps}
\caption{
Degree-dependent mean intervertex separation in stochastic networks (trees) growing under the mechanism of random attachment. These networks have exponential degree distributions. 
(a) At each time step, a vertex is attached to a randomly chosen vertex of the network. The dependence is the result of the simulation of the network of $10^5$ vertices, $50$ runs. For comparison, a line with a slope $-1/2$ is shown.  
(b) At each time step, 3 vertices are attached to a randomly chosen vertex of the network. The dependence is presented for the network of $9998$ vertices, $50$ runs. 
The initial configuration consists of two vertices connected by an edge. 
For comparison, a line with a slope $-1/4$ is shown. 
Note that in these plots $\max \ell(k) \approx 2\min  \ell(k)$. In other words, in these networks, there are no vertices of degree greater than $k_\text{max}$: $\ell(k_\text{max})=\max \ell(k)/2$. Note fluctuations in the range of the highest degrees.  
}
\label{f2}
\end{figure}

Formula (\ref{e1}) fails at $\gamma \to \infty$. E.g., it cannot be applied for networks with an exponential degree distribution. In growing trees with this distribution, 
we observe the dependence: 
%%%%%the resulting dependence turned out to be   
%%
\begin{equation} 
\ell(k) \cong A \ln N - C k
\, , 
\label{e2}
\end{equation}  
where the constants $A$ and $C$ depend on a network. 
In particular, we found that this law is exact  
%%%%%We obtained this dependence analytically for 
in deterministic graphs (trees) with an exponential degree distribution [e.g., graph (e) in Fig.~\ref{f1}] at least up to very large sizes. 
Moreover, we observed the same dependence in a 
simulated 
stochastically growing tree with random attachment. 
In this tree (with an exponential degree distribution), at each time step, a new vertex is attached to a randomly selected vertex of the net. The result of the simulation of this network is shown in Fig.~\ref{f2}(a). In both the networks---graph (e) in Fig.~\ref{f1} and the corresponding stochastic net with random attachment---the slope of the degree dependence turned out to be $-1/2$. 
More generally, if in a growing tree of this kind, at each step, $n$ new vertices become attached to a vertex, the slope of the degree dependence equals $-1/(n+1)$ [see Fig.~\ref{f2}(b)]. 

%%%%%%%%%%%%%%%%%%%%%%%%%%%%%%%%%%%%%%%%%%%%%%%%%%%%%%%%%%%%%%%%%%%%
%%%%%%%%%%%%%%%%%%%%%%%%%%%%%%%%%%%%%%%%%%%%%%%%%%%%%%%%%%%%%%%%%%%%

%%%%%%%%%%%%%%%%%%%%%%%%%%%%%%%%%%%%%%%%%%%%%%%%%%%%%%%%%%%%%%%%%%%%
%%%%%%%%%%%%%%%%%%%%%%%%%%%%%%%%%%%%%%%%%%%%%%%%%%%%%%%%%%%%%%%%%%%%

All networks that we studied, had the generic property: 
\begin{equation} 
\max_k\ell(k) 
\approx 
%%= 
2\min_k\ell(k)
\, , 
\label{e3}
\end{equation} 
in the large network limit. As is natural, the maximum value of $\ell(k)$ is attained at the minimal degree of a vertex in a network, and vise versa, the minimum value of $\ell(k)$ is attained at the maximum degree.

\section{$\ell(k)$ of an uncorrelated network}\label{s-uncorrelated}

The configuration model \cite{b80,bbk72,bc78,w81} is a standard model of an uncorrelated (equilibrium) random network. In simple terms, these are maximally random graphs with a given degree distribution. In the large network limit, they have relatively few loops and almost surely are trees in any local environment of a given vertex. 
The mean intervertex distance $\ell$ in these networks is estimated in the following way, Ref.~\cite{nsw02} (see also Refs.~\cite{dms03,ffh02}). 
The mean number of $m$-th nearest neighbors of a vertex is 
\begin{equation} 
z_m = z_1 (z_2/z_1)^{m-1}
\, , 
\label{e4}
\end{equation}
where $z_1 = \langle k \rangle$ is the mean number of the nearest neighbors of a vertex, i.e., the mean degree. $z_2 = \langle k^2 \rangle - \langle k \rangle$ is the mean number of the second nearest neighbors of a vertex. 
$z_2/z_1$ is actually the branching coefficient. By using formula (\ref{e4}), one can get $\ell$: $z_\ell \sim N$, 
so $\ell(N) \approx \ln N/\ln(z_2/z_1)$. 

Similarly, for the mean number of $m$-th nearest neighbors of a vertex of degree $k$, we have 
\begin{equation} 
z_m(k) = k (z_2/z_1)^{m-1}
\, . 
\label{e5}
\end{equation}
So, the estimate is $k (z_2/z_1)^{\ell(k)-1} \sim N$ and thus 
\begin{equation} 
\ell(k) \approx \frac{\ln (N/k)}{\ln(z_2/z_1)}
\, . 
\label{e6}
\end{equation} 
Here we neglected an additional constant independent of $N$ and $k$ which would be excess precision.   

The relation (\ref{e5}) is evident. It also may be obtained strictly by using the $Z$-transformation technique: 
\begin{equation} 
z_m(k) = \left[x\frac{d}{dx}\phi^k_1(\phi_1(\ldots\phi_1(x)))\right]_{x=1}
\, . 
\label{e7}
\end{equation} 
$\phi_1(x)=\phi(x)/z_1$ is the $Z$-transformation of the distribution of the number of edges of an end vertex of an edge with excluded edge itself. $\phi(x)$ is the $Z$-transformation of the degree distribution of the network: 
$\phi(x) \equiv \sum_k P(k) x^k$. 
Formula (\ref{e7}) is a direct consequence of the following features of the configuration model: (i) the network has a locally tree-like structure, (ii) vertices of the network are statistically equivalent, (iii) correlations between degrees of the nearest neighbor vertices are absent.  
Relation (\ref{e7}) together with $\phi_1(1)=\phi(1)=1$ readily leads to relation (\ref{e6}). 

Note that expression (\ref{e6}) also follows from the mentioned result of Holyst {\em et al.}, Ref.~\cite{ffh04}, that is $\ell(k,k') \approx \ln[N/(kk')]/\ln(z_2/z_1)$ for the configuration model. Substituting this result into formula (\ref{c1}) and ignoring terms independent of $N$ and $k$ immediately gives expression (\ref{e6}). In its turn, substituting expression (\ref{e6}) into formula (\ref{c1}) leads to a standard formula for the configuration model: 
$\ell \approx \ln N / \ln(z_2/z_1)$.   

One point should be emphasized. In the configuration model, the logarithmic size dependence of the (degree-independent) mean intervertex distance $\ell(N) \sim \ln N$ is valid only for degree distributions with a finite second moment $\langle k^2 \rangle$. If $\langle k^2 \rangle$ diverges as $N \to \infty$, $\ell(N)$ grows slower than $\ln N$. One can see that the result (\ref{e6}) may be generalized to any given form $\ell(N)$ of the size-dependence of the mean intervertex distance.  
In this general case, the degree-dependent separation is expressed in terms of the function $\ell(N)$, namely,    
$\ell(k,N) \sim \ell(N/k)$.

\section{Derivations}\label{s-derivations}

In this section we 
study  
%%%%%obtain 
a degree-dependent intervertex separation in the deterministic graphs of Fig.~\ref{f1}. Graphs (a) -- (d) have a discrete spectrum of vertex degrees with a power-law envelope. Graph (e) has a discrete spectrum of vertex degrees with an exponential envelope. We also list some basic characteristics of these graphs. We stress that the main structural characteristics (clustering, degree--degree correlations \cite{kr01,pvv01,vpv02,ms02,msz03,n02}, etc.) of these deterministic networks are quite close to those of their stochastic analogs (see \cite{dgm02}). 

{\em (A) Graph (a) in Fig.~\ref{f1}.}---This graph was proposed in Ref.~\cite{dm02} and extensively studied in Ref.~\cite{dgm02}. The growth starts from a single edge ($t=0$). 
At each time step, each edge of the graph transforms into a triangle. 
Actually, we have a deterministic version of a stochastic growing network with attachment of a new vertex to a randomly chosen edge, see Ref.~\cite{dms01}. 
The number of vertices of the graph is $N_t=1 + (3^{t}+1)/2$. ($t=0,1,2,\ldots$ is the number of the generation.) In the large network limit, the mean degree of the graph is $\langle k \rangle \to 4$. 

Degrees of the vertices in the graph take values $k(s) = 2^s$, $s=1,2,\ldots,t$. The spectrum of degrees has a power-law envelope. This spectrum corresponds to a continuum scale-free spectrum $P(k) \propto k^{-\gamma}$ with exponent 
$\gamma = 1+\ln3/\ln2 = 2.585\ldots$. Note that this network has numerous triangles, which suggests high clustering. In more detail, by definition, 
the average clustering coefficient of a vertex of degree $k$ is  
\begin{equation} 
C(k) = \left\langle\frac{c(k)}{k(k-1)/2}\right\rangle_k = 
\frac{\langle c(k)\rangle_k}{k(k-1)/2}
\, . 
\label{e8}
\end{equation} 
Here, $c(k)$ is the number of triangles attached to a vertex of degree $k$, and $\langle \ \rangle_k$ means the averaging over all vertices of 
degree $k$. One can see that in this graph (as well as in its stochastic version)  
\begin{equation} 
C(k) = \frac{2}{k} 
\, . 
\label{e9}
\end{equation} 
[Indeed, by construction, the number of triangles attached to a vertex of degree $k$ in the graph is $k-1$. So, $C(k) = (k-1)/[k(k-1)/2]=2/k$.]  
This gives, for the mean clustering,  
\begin{equation} 
\overline{C} = \sum_k P(k)C(k) = \frac{4}{5}
\, ,  
\label{e10}
\end{equation} 
while the standard clustering coefficient (transitivity), i.e., the density of loops of length $3$ in a network, 
\begin{equation} 
C = \frac{\sum_k P(k)C(k)k(k-1)}{\sum_k P(k)k(k-1)}
\, ,  
\label{e11}
\end{equation} 
approaches zero in the infinite network limit, $C=0$. 
Note the difference between the finite mean clustering of the network and its zero clustering coefficient. 

%%*************************************
%%
%%(1) PSEUDOFRACTAL 
%%
%%$\gamma = 1+\ln3/\ln2 = 2.585\ldots$  
%%
%%$\overline{k} \to 4$
%%
%%STARTS WITH $t=0$ -- ONE EDGE
%%
%%$N_t=1 + (3^{t}+1)/2$ 
%%
%%$k(s) = 2^s$, $s=1,2,\ldots$ 
%%
%%$$
%%C(k) = \left\langle\frac{c(k)}{k(k-1)/2}\right\rangle_k = 
%%\frac{\langle c(k)\rangle_k}{k(k-1)/2}
%%$$
%%
%%$$
%%C(k) = \frac{2}{k}
%%$$ 
%%
%%$$
%%\overline{C} = \sum_k P(k)C(k)
%%$$
%%
%%$\overline{C} = 4/5$
%%
%%$$
%%C = \frac{\sum_k P(k)C(k)k(k-1)}{\sum_k P(k)k(k-1)}
%%$$
%%
%%$C \to 0$ as $N \to \infty$ 

In principle, one may derive an exact analytical expression for the degree-dependent separation by using recursion relations and the $Z$-transformation technique. 
However, these calculations turn out to be cumbersome. 
Instead, here we only check that some analytical formula for $\ell(k)$ is valid in a sufficiently large number of generations of a deterministic graph, up to, say, $t \sim 10$ or $12$. So, we confirm a guessed expression in networks of sizes up to $N \sim 10^5$.  
%%However, here we demonstrate a more rapid way which turns out to be useful in many situations: 
In fact, we implement the following approach: 

\begin{itemize}

\item[(i)] 
Find the  mean separation values $\ell_t(s)$ for all kinds of vertices in each of several first generations of the deterministic graph [$t$ is the number of generation, and $k=2^s$, $s=1,2,\ldots,t$]; 

\item[(ii)] 
by using this array of numbers, guess the form of $\ell_t(s)$; 

\item[(iii)] 
check this result by computing directly $\ell_t(s)$ for several extra generations of the graph. 

\end{itemize}
There are few computations in stage (i): we have to find only $t$ values of $\ell_t(s)$ in a $t$ generation of a graph. For sufficiently small networks, these values can be found even without a computer. 
Step (ii) also turns out to be rather easy since we already know the structure of the analytical expressions for a mean intervertex distance in these networks (see Ref.~\cite{dgm02}). 
Step (iii) may be performed by using a computer to count paths. 
%%%%%One should stress that this effective approach, nevertheless, is somewhat ``unscientific'': It 
This approach is based on our experience with problems on these graphs and was checked in Ref.~\cite{dgm02} for related quantities. Our guess actually exploits underlined recursion relations without revealing them. 
%%%%%A mathematician would prove that the assumed form is exact by applying induction. In contrast, we, physicists, only check that this form exactly works in a few higher generations. Thus, staying on the intuitive level, we do not prove our assumption but only demonstrate that it is valid in large enough graphs and further assume that it will work even in graphs of higher generations. 
Nonetheless, we can only claim that the analytical expressions, obtained in this way, are valid at the studied generations of our deterministic graphs. In principle, there exists a (small) chance that at some higher generation (or generations), these formulas fail. Thus, the results of this section should be considered only as observations of $\ell(k)$ for a set of networks of a modest size.  
   
In this way, we get 
\begin{equation} 
\ell_t(s) = \frac{1}{2(N_t-1)}[2(2t-s+5)3^{t-2} - 3^{s-1} + 1]
\, .  
\label{e12}
\end{equation}  
This 
%%%%%exact result 
formula is valid for $t \geq 1$. We checked it up to $t=12$, which corresponds to $N_t = 265\,722$. We also checked that this formula leads to the known exact formula for the mean intervertex distance $\ell$ for any $t$ and so that $N$ \cite{dgm02}.   
An asymptotic form of this expression is 
\begin{equation} 
\ell(k,N) = \frac{4}{9\ln3}\ln N - \frac{2}{9\ln2}
%%(1+\frac{2}{N})
\ln k - \frac{k^{\gamma-1}}{6N} + 
\frac{4}{9}\,\frac{\ln2}{\ln3} + \frac{10}{9}
+ \ldots
\,   
\label{e13}
\end{equation} 
at large $N$, where $N$ is the total number of vertices in the graph. 
%%%%%%%%%%%%%%%%%%%at large $k$ [$k \gg (\ln N)^{1/(\gamma-1)}$, $N$ is the total number of vertices in the graph]. 
This leads to 
%%%%result 
formula (\ref{e1}). 

One can see that the minimum value of $\ell(k)$ is 
$\ell_\text{min}=\ell(k=2^t) \cong 2t/9$, where $t \cong \ln N/\ln3$. 
On the other hand, its maximum value is 
$\ell_\text{max}=\ell(k=2) \cong 4t/9$. 
So, we arrive at relation (\ref{e3}): $\ell_\text{max}=2\ell_\text{min}$. 
 
%%
%%$k_\text{min}=2$ at $s=1$ 
%%
%%$$
%%\ell_\text{max}=\ell(k=2) \cong 4t/9
%%$$
%%where $t \sim \ln N/\ln3$. 
%%
%%$k_\text{max}=2^t$ at $s=t$ 
%%
%%$$
%%\ell_\text{min}=\ell(k=2^t) \cong 2t/9
%%$$

{\em (B) Graph (b) in Fig.~\ref{f1}.}---This graph was proposed in Ref.~\cite{jkk02}. At each time step, each edge of the graph transforms in the following way: each end vertex of the edge gets a new vertex attached [see Fig.~\ref{f1}, graph (b), instant 0 $\to$ instant 1]. This graph is very similar to graph (a). In particular, the exponent of its degree distribution is the same,  
$\gamma = 1+\ln3/\ln2 = 2.585\ldots$. The difference is that the graph is a tree, so the mean degree $\langle k\rangle \to 2$ as $N \to \infty$. 

The total number of vertices in the graph is $N_t = 3^t+1$. The vertices have degrees $k(s) = 2^s$, where $s=0,1,2,\ldots,t$. In the same way as for graph (a), we find the expression
\begin{equation} 
\ell_t(s) = \frac{1}{2(N_t-1)}[(4t-2s+9)3^{t-1} - 3^s]
\, ,   
\label{e14}
\end{equation} 
which is 
%%%%%%valid 
observed starting with $t=0$. 
This leads to the asymptotic relation 
\begin{equation} 
\ell(k,N) = \frac{2}{3\ln3}\ln N - \frac{1}{3\ln2}\ln k 
- \frac{k^{\gamma-1}}{2N} + 
\frac{3}{2} 
+ \ldots
\, ,   
\label{e15}
\end{equation} 
that is, to formula (\ref{e1}). 

The minimum value of $\ell(k)$ is $\ell_\text{min}=\ell(k=2^t) \cong t/3$, where $t \cong \ln N/\ln3$. The maximum value is $\ell_\text{max}=\ell(k=1) \cong 2t/3$, i.e., again, we arrive at relation (\ref{e3}). 

%%(2) AT EACH STEP EACH EDGE CREATES 2 EXTRA INDEPENDENT EDGES -- TREE 
%%
%%STARTS WITH $t=0$ -- ONE EDGE 
%%
%%$\overline{k} \to 2$
%%
%%$\gamma = 1+\ln3/\ln2 = 2.585\ldots$ 
%%
%%$k(s) = 2^s$, $s=0,1,2,3,\ldots$ 
%%
%%$N_t = 3^t+1$
%%
%%$$
%%\ell_t(s) = \frac{1}{2(N_t-1)}[(4t-2s+9)3^{t-1} - 3^s]
%%$$
%%
%%-- valid starting with $t=0$ ($s=0$) 
%%
%%$$
%%\ell(k,N) = \frac{2}{3\ln3}\ln N - \frac{1}{3\ln2}\ln k 
%%- \frac{k^{\gamma-1}}{2N} + 
%%\frac{3}{2} 
%%+ \ldots
%%$$ 
%%
%%$k_\text{max}=2^t$ at $s=t$ 
%%
%%$$
%%\ell_\text{min}=\ell(k=2^t) \cong t/3
%%$$
%%
%%$k_\text{min}=1$ at $s=0$ 
%%
%%$$
%%\ell_\text{max}=\ell(k=1) \cong 2t/3
%%$$
%%where $t \sim \ln N/\ln3$. 

{\em (C) Graph (c) in Fig.~\ref{f1}.}---At each step, (i) a new vertex becomes attached to each end vertex of each edge of this graph and, simultaneously, (ii) a new vertex becomes attached to each vertex of the graph. This produces a growing deterministic scale-free tree with exponent $\gamma=3$, which is a deterministic analog of the Barab\'asi-Albert model \cite{ba99,baj99} (for exact solution of the stochastic model, see Refs.~\cite{dms00,kr01,krl00}). 

%%(3) AT EACH STEP EACH EDGE CREATES 2 EXTRA INDEPENDENT EDGES PLUS EACH VERTEX RECEIVES ONE EDGE -- TREE 
%%
%%STARTS WITH $t=0$ -- ONE EDGE
%%
%%$\overline{k} \to 2$

The number of vertices in the graph is 
$N_t = 1+ (4^{t+1}-1)/3$. Their degrees take values 
$k(s)=2^s-1$, $s=1,2,3,...,t+1$. 
The 
%%%%%%%resulting formula for the 
observed degree-dependent separation is   
\begin{equation} 
\ell_t(s\geq2) = \frac{1}{9(N_t-1)}[2(6t-3s+10)4^t - 4^s - 1]
\, .    
\label{e16}
\end{equation} 
%%valid for $t \geq 0$, $s \geq 1$
Asymptotically, this is 
\begin{equation}  
\ell(k,N) = \frac{1}{\ln4}\ln N - \frac{1}{2\ln2}\ln k 
- \frac{k^{\gamma-1}}{9N} + 
\frac{\ln3}{2\ln2} + \frac{2}{3} 
+ \ldots
\, 
%.    
\label{e17}
\end{equation} 
for 
%%%%%%%%%%%%%%%%%$k^3 \gg N$ 
$k,N \gg 1$ (note that the maximum degree of a vertex in this graph is $k_\text{max} \sim N^{1/2}$). 
This leads to expression (\ref{e1}) with $\gamma=3$, which coincides 
with result (\ref{e6}) for uncorrelated networks. This is an understandable coincidence. Indeed, correlations between degrees of the nearest neighbor vertices in this deterministic graph, as well as in the Barab\'asi-Albert model are anomalously week. So, the result must be close to that for an uncorrelated network. 

%%$k_\text{max}=2^{t+1}-1$ corresponds to $s=t+1$ 

The minimum value of $\ell(k)$ in this graph is 
$\ell_\text{min}=\ell(k=2^{t+1}-1) \cong t/2$, where $t \sim \ln N/\ln4$. 
The maximum value is 
$\ell_\text{max}=\ell(k=1) \cong t$, 
so that relation (\ref{e3}) is fulfilled. 

%%
%%$k_\text{min}=1$ at $s=1$ 
%%
%%$$
%%\ell_\text{max}=\ell(k=1) \cong t
%%$$
%%where $t \sim \ln N/\ln4$. 

{\em (D) Graph (d) in Fig.~\ref{f1}.}---At each step, (i) a pair of new vertices is attached to ends of each edge of the graph plus (ii) two new vertices are attached to each vertex of the graph. This results in the value of the $\gamma$ exponent greater than $3$, $\gamma = 1+\ln5/\ln2 = 3.322\ldots$. 

%%(4) EACH EDGE CREATES 2 EXTRA INDEPENDENT EDGES PLUS EACH VERTEX RECEIVES 2 EDGEs -- TREE 
%%
%%STARTS WITH $t=0$ -- ONE EDGE
%%
%%$\overline{k} \to 2$
%%
%%$\gamma = 1+\ln5/\ln2 = 3.322\ldots$ 

The number of vertices in the graph is 
$N_t = (3 \cdot 5^t + 1)/2$. 
Degrees of the vertices are 
$k(s) = 3 \cdot 2^{s-1}-2$, $s=1,2,3,...,t+1$.  
The observed expression for the degree-dependent separation is 
\begin{equation} 
\ell_t(s) = \frac{1}{8(N_t-1)}[(72t-36s+71+5^{3-s})5^{t-1} + 2\,5^{s-1} - 6]
\, .     
\label{e18}
\end{equation}
The corresponding asymptotic expression is of the following form:  
\begin{equation} 
\ell(k,N) = \frac{6\ln N}{5\ln5}
%%\ln N 
- \frac{3\ln k }{5\ln2}
%%\ln k 
- \frac{5^{-\ln3/\ln2}}{4N}k^{\gamma-1} + 
1.232
%%\frac{6\ln(2/3)}{5\ln5} + \frac{3\ln(3)}{5\ln2} + \frac{7}{12} 
+ \ldots
\,,     
\label{e19}
\end{equation} 
where the contribution 
$1.232\ldots = [6\ln(2/3)]/(5\ln5) + (3\ln3)/(5\ln2) + 7/12$. 
Again, now with the graph where $\gamma>3$, we arrive at formula (\ref{e1}). 

In this graph, we have $\ell_\text{min}=\ell(k=3 \cdot 2^t-2) \cong 3t/5$ and 
$\ell_\text{max}=\ell(k=1) \cong 6t/5$, where $t \cong \ln N/\ln5$. 

%%$k_\text{max}=3\,2^t-2$ corresponds to $s=t+1$ 
%%
%%$k_\text{min}=1$ at $s=1$ 
%%
%%
%%Note that we have all the time 
%%
%%$$
%%\text{const}\,\ln[N/k^{(\gamma-1)/2}]
%%$$, 
%%the main contribution. This allows us to hope that the case $\gamma = \infty$ (i.e., an exponential degree distribution) will be very different. Indeed, we see
%%below... 
%%
%%Note that in the case $\gamma=3$ we have anomalously week degree-degree correlations in growing networks (practically absent), we have the main contribution 
%%
%%$\ell(k) \cong \text{const}\,\ln (N/k)$ 
%%
%%precisely as it is in the configuration model

The important feature of the expressions for $\ell(k,N)$ in deterministic scale-free networks with $\gamma\neq 3$ were non-equal coefficients of $\ln N$ and $\ln k$. For comparison we have measured $\ell(k,N)$ in a random growing scale-free network growing through the mechanism of preferential attachment with a linear preference function \cite{dms00}. At each time step, a new vertex emerges and becomes attached to a vertex chosen with probability proportional to the sum of its degree and a constant $A$. Exponent $\gamma = 3+A$. We use $A=1$, so that $\gamma=4$. The resulting degree-dependent separations are shown in Fig.~\ref{f1prime}(a) for networks of up to 300\,000 vertices. One can see in the inset that in these random networks, the difference $\ell(k=1,N)-\ell(k,N)$ is independent of $N$ in contrast to the deterministic graphs (a)---(d). Furthermore, $[\ell(k=1,N)-\ell(k,N)]/\log_{10} k \approx 1.0$ 
%%if $\ln k$ is small enough. 
as $\log_{10} k$ approaches zero  
[i.e., $d\ell(k,N)/d\ln k \approx -0.43$]. 
However, at large $k$, we find a linear dependence on $\log_{10} k$ with a larger slope, namely $4.1$ [i.e., $d\ell(k,N)/d\ln k \approx -1.8$].  
In its turn, $\ell(k=1,N)$ is well fitted by a linear dependence on $\log_{10} N$ with a slope approximately $3.1$, see Fig.~\ref{f1prime}(b) [i.e., 
$d\ell(k=1,N) /d\ln N \approx 1.35$]. The 
%%%%%%%%%%%strong 
difference in these slopes --- $4.1$ and $3.1$ --- is in sharp contrast to uncorrelated networks. 
%%%%%%%%%%%Of course, these particular numbers could not be predicted by using deterministic graphs. 
%%Surprisingly, the 
The ratio of these slopes, $1.3$ is close to what we had for deterministic graphs according to Eq.~(\ref{e1}) with $\gamma=4$ substituted, namely, $(\gamma-1)/2=1.5$. 
%%On the other hand, the ratio of these numbers differs from the corresponding ratios for deterministic graphs. 
%%%%%%%%%%%Nonetheless, 
Moreover, Fig.~\ref{f1prime}(a) shows that for each network size, 
$\ell_\text{max} \approx 2\ell_\text{min}$, as was observed in deterministic graphs.  

One should note that the contribution $\sim k^{\gamma-1}/N$ to $\ell(k,N)$ for the deterministic graphs, is noticeable only in a narrow neighborhood of $k_{\max}$, if results are presented in the form $\ell(k,N)$ vs. $\ln k$. On the other hand, the linear dependence $\ell(k,N)$ on $\ln k$ is realized in a much wider range of $\ln k$. In Eq.~(\ref{e13})---graph (a), it is valid for all degrees up to nearly $k_{\max}$, and in Eqs.~(\ref{e15}), (\ref{e17}), and (\ref{e19})---graphs (b), (c), and (d), respectively, this law is observable for $k \gg 1$. It is in this region that we compared the rations of the coefficients of $\ln k$ and $\ln N$ in deterministic and stochastic growing scale-free networks.  

{\em (E) Graph (e) in Fig.~\ref{f1}.}---At each time step, a new vertex becomes attached to each vertex of the graph. The growth starts with a single vertex ($t=-1$). The total number of vertices in the graph is $N_t = 2^{t+1}$. 
The degree distribution is exponential. One can check that the number of vertices of degree $k$ at time $t$ is $N_t(k\leq t) = 2^{t+1-k}$, 
$N_t(k = t+1) = 2$ ($t$ is assumed to be greater than $-1$). 

By using the above described procedure, we find the exact expression: 
%%
%%(5) EXPONENTIAL 
%%
%%STARTS WITH $t=0$ -- ONE EDGE - AT EACH STEP EACH VERTEX RECEIVES ONE EDGE
%%
%%
\begin{equation} 
\ell_t(k) = \frac{2^t}{2^{t+1}-1}\,(2t+2-k)
\, .     
\label{e20}
\end{equation}
%%is valid even at $t=-1$, $k=0$ -- a single vertex. 
This formula shows that the linear dependence on degree is valid for any $k$. 
For the large graphs we have
\begin{equation}
\ell(k,N) \cong \frac{\ln N}{\ln 2} - \frac{k}{2} 
\, ,      
\label{e21}
\end{equation}
which confirms formula (\ref{e2}). 

%%The maximum degree of a vertex in the graph is $k_{\text{max}}=t+1=\ln N/\ln2$, and so 
In this graph, $\ell_\text{min} \cong \ln N/(2\ln2) \cong \ell_\text{max}/2$ which coincides with relation (\ref{e3}).
%%$\ell(k)$ decreases from 
%%$\ell_\text{max} = \ell(k=1) = (2t-1)2^t/(2^{t+1}-1) \cong \ln N/\ln2-1/2$ 
%%to 
%%$\ell(k_{\text{min}}) = (t+1)2^t/(2^{t+1}-1) \cong \ln N/(2\ln2)$
%%

Graph~(e) has a close stochastic analog---a tree, where at each step, a new vertex is attached to a randomly chosen vertex. 
%%%%%In principle, this is a solvable model. 
%%Nonetheless, it is very 
It is easy to obtain the asymptotic expression for the mean shortest path length $\ell(N)$ in this network. Let us consider even more general model. Let at each time step, $n$ new vertices be attached to a randomly selected vertex. Then the total number of vertices $N$ grows as $N_t \cong nt$. For the total length of the shortest paths between vertices in the network at time $t+1$ one can wright: 
\begin{eqnarray}
\!\!\!\!\!\!\!\!\!\!\!\!\!\!\!\!\!\!&& 
\frac{N_{t+1}(N_{t+1}-1)}{2}\,\ell(t+1) = \frac{N_t(N_t-1)}{2}\,\ell(t)
\nonumber 
\\[5pt]
\!\!\!\!\!\!\!\!\!\!\!\!\!\!\!\!\!\!&& 
+ \frac{1}{N_t}\, N_t \Biggl(\! 1\cdot n + 2\,\frac{n(n-1)}{2} + 
n(N_t - 1)[\ell(t) + 1] \!\Biggr)
%%\, 
.      
\label{c5}
\end{eqnarray}
The first term on the right-hand side of this equation is the total length of the shortest paths in the network at time $t$. The second term is the increase of this total length due to the attachment of $n$ new vertices to a randomly chosen vertex. The factor $1/N_t$ is due to the random choice. The term $1\cdot n$ is the sum of the paths connecting the new vertices to their ``host''. 
The term $2\cdot n(n-1)/2$ is the total length of the paths between the new vertices. 
The last term in the large parentheses is the sum of the lengths of the paths connecting the $n$ new vertices and the $N_t-1$ old vertices distinct from the vertex receiving new connections.    
In the large network limit, Eq.~(\ref{c5}) is readily reduced to the following one: 
\begin{equation}  
\frac{N^2}{2} n \frac{d\ell}{dN} = - \frac{n(n+1)}{2}\ell + nN \cong nN 
,      
\label{c6}
\end{equation}
and so we have 
\begin{equation}
\ell \cong 2 \ln N 
,      
\label{c7}
\end{equation}
independently of $n$. 

The calculation of $\ell(k)$ is a more difficult problem.   
So, for comparison, we present here only the result of the simulation of this stochastic network. Figure~\ref{f2}(a) demonstrates that the dependence $\ell(k)$ in the stochastically growing network is a linear function with the same slope $-1/2$ as in the deterministic small world (e) in Fig.~\ref{f1}. 

We also considered more general deterministic graphs of this type, where $n$ new vertices become attached to each vertex of a network at each time step. 
The resulting dependence $\ell(k)$ is a linear function but with slope $-1/(n+1)$. Figure~\ref{f2}(b) shows that $\ell(k)$ of the corresponding stochastically growing networks has the same form. We also checked that $\ell(k=1,N) \approx 2 \ln N$, as in expression (\ref{c7}) for $\ell(N)$.

\section{Discussion and summary}\label{s-discussion}

Several points should be emphasized:

(i) One can estimate a typical value of the correction term in formula (\ref{e1}). At the maximum degree $k_{\text{max}} \sim N^{1/(\gamma-1)}$, this term is of the order of $k_{\text{max}}^{\gamma-1}/N \sim \text{const}$. This should be compared to $\ln[k_{\text{max}}^{1/(\gamma-1)} \sim \ln N]$. 

%%%%%(iii) We believe that the correction term in formula (\ref{e1}) can be hardly observed due to its smallness. It is the form of the main contributions in formulas (\ref{e1}) and (\ref{e2}), that is important. We found that these expressions differ from that for the configuration model, formula (\ref{e6}). 

%%%%%(iv) 
(ii) 
One should indicate that law (\ref{e2}), i.e., a linear dependence $\ell(k)$, was obtained only for growing trees with an exponential degree distribution. In non-tree growing networks with random attachment (at each time step, a new vertex becomes attached to several randomly chosen vertices), we observed a non-linear dependence. 

%%%%%(v) 
(iii) 
The relative width of the distribution of the intervertex distance in infinite small worlds approaches zero \cite{dms03,dgm02}. In other words, vertices of an infinite small world are almost surely mutually equidistant.  
This circumstance does not allow one to measure $\ell(k)$ in an infinite network with the small-world effect. 
However, even in very large real-world networks (e.g., in the Internet \cite{vpv02}), the distribution of the intervertex distance is still broad enough. 
So, in real networks, $\ell(k)$ is a measurable characteristic. 

%%%%%(vi) 
(iv) 
The degree-dependent mean intervertex distance may be considered as a measure of ``centrality'' of a given degree vertex in a network. How does this characteristic relate to other centrality characteristics \cite{n90}, first of all to the centrality index of a vertex \cite{s66}? Recall that the centrality index of a vertex $v$ is defined as $c_v = (N-1)/\sum_{u}\ell(v,u)$, where $\ell(v,u)$ is the length of the shortest path between vertices $u$ and $v$, $N$ is the number of vertices in the graph, and the sum is over all vertices of the graph. (The centrality index is often given without the $N-1$ factor.) One may see that the mean centrality index $c(k)$ of a vertex of degree $k$ is related (but not equal) to $1/\ell(k)$. Nevertheless, there is a special case---graphs where 
every vertex of a given degree $k$ has the same value of the sum of intervertex distances between this and the rest of the vertices. So, this value is exactly $(N-1)\ell(k)$, and consequently $c(k)=1/\ell(k)$.  
%%all vertices of a given degree $k$ have the same value of the sum of intervertex distances. 
%%In these graphs, 
%%$c(k)=1/\ell(k)$. It is the case 
This situation is realized in our deterministic graphs. Thus, in the deterministic graphs, we actually found the inverse centrality index, but in random networks, $c(k)$ and $\ell(k)$ are different characteristics.

In conclusion, we have 
%%%%%found 
studied the mean length of the shortest paths between a vertex of degree $k$ and the other vertices in 
%%%%%a number of 
growing 
networks with power-law and exponential degree distributions. 
In the investigated deterministic and random networks, we  
%%%%%
%%%%%We have obtained these dependences by using a representative set of deterministic graphs. We have checked that these laws are also realized at least in several stochastically growing networks. 
%%%%%We 
have observed dependences $\ell(k)$ which strongly differ from those for uncorrelated networks. 
%%So, w
%%%%%We believe that our results hold for a wide class of random networks. 
Our results characterize the compactness of a network from the point of view of a vertex with a given number of connections.

\begin{acknowledgments} 

This work was partially supported by projects 
POCTI/FAT/46241/2002\,and\,POCTI/MAT/46176/2002. 
SND and JFFM acknowledge the NATO program OUTREACH for support. JGO
acknowledges financial support of FCT (Portugal), grant No. SFRH/BD/14168/2003.
Authors thank A.N.~Samukhin for useful discussions.

\end{acknowledgments}

\end{document}